\documentclass{aa} 
\usepackage{psfig}
\def\h50{h_{50}^{-1}}              %
\def\lya{Ly$\alpha$}
\def\mr{m_\mathrm{R}}              %
\def\Mr{M_\mathrm{R}}              %
\begin{document}
\thesaurus {03(11.17.1; 11.08.1; 11.09.4)} 
\title{Detection of candidate primordial galaxies at $z>4$
\thanks{Based on observations made with the ESO 3.6m Telescope}
}
\author{V. Le~Brun \inst{1}, J. Boulesteix \inst{2}, A. Mazure \inst{1}, P. 
Amram \inst{2}, G. Court\`es \inst{2}}
\institute{IGRAP, Laboratoire d'Astronomie Spatiale du C.N.R.S., B.P. 8, 
F-13376 Marseille CEDEX 12, France, lebrun@lasa13.astrsp-mrs.fr, 
mazure@astrsp-mrs.fr
\and IGRAP, Observatoire de Marseille, 2 Place Le Verrier, 
F-13248 Marseille CEDEX 04, France, boulesteix@obmara.cnrs-mrs.fr, 
amram@obmara.cnrs-mrs.fr
}
\offprints{V. Le~Brun}
\date {Received November 1997; accepted ?}
\maketitle
\markboth{V. Le~Brun et al.: Primordial galaxies at $z>4$}{}
\begin{abstract}
We report on the detection of four primordial galaxies candidates in the 
redshift range $4.55 \le z \le  4.76$, as well as 4 other candidates in the 
range $4.12 \le z \le 4.32$. These galaxies 
have been detected with the now common technique of the Lyman break, but with
a very original instrumental setup, which allows the simultaneous detection
on a single exposure of three narrow ($\mathrm{FWHM} \sim 200$~\AA)
bands. Depending on between which pair of bands the Lyman break is located,
it thus allows to detect two ranges of redshifts. 
Two of the the higher redshift candidates are located only $50\h50$~kpc
apart, thus forming a possible primordial galaxy pair. 
\end{abstract}   
%
\section{\label{intro}Introduction}
The knowledge of the epoch when galaxies formed, that is when they 
formed the bulk 
of their stars, is a major cosmological question. The detection of such 
primordial galaxies 
would give constraints on their formation, their chemical evolution or the 
formation of large scale
structures. All the attempts to detect these galaxies at redshifts 
$z\ge3$ by 
detecting the Ly$\alpha$ emission line, either by 
visible and IR spectroscopy (Koo \& Kron 1980, Schneider et al. 1991, Thompson \& Djorgovski
1995), narrow band imaging (De Propris et al. 1993, Parkes et al. 1994)
or Perot-Fabry (Thompson et al. 1995) have failed, while the Ly$\alpha$ 
flux of a $z=4$
galaxy, $\sim 10^{-16\pm 1}$~erg~cm$^{-2}$~s$^{-1}$ (Thompson \& Djorgovski 1995), should have been 
easily observed with 3.6m
telescopes. Only recently, Hu (1998) reported the detection of \lya\ emitters
at redshifts as high as 4.5. However, it is now known that even small amounts 
of dust can 
absorb the great majority
of the Ly$\alpha$ photons by resonant scattering (Charlot \& Fall 1993).

The most promising method to observe these galaxies seems to use the Lyman 
break at 912~\AA. This
break is present in all kinds of galaxies, for all reasonable stellar formation 
scenarii (Bruzual \& Charlot 1993), and 
is observed in the visible range for $z\ge 3.5$. 

The first very high redshift galaxies have been observed through multi band 
imaging, with broad filters
located on each side of the Lyman break at $\left<z\right>=3.2$, selected using 
color criteria, and several of the candidates have recently been spectroscopically 
confirmed with the Keck telescope (Steidel \& Hamilton 1992,1993, Steidel et al. 
1995, Pettini et al. 1997). 
The average projected density of these objects is expected to be 
0.5~gal~arcmin$^{-2}$ at $R \sim 25$ and $\left<z\right> = 3.2$ (Steidel et
al. 1995).

In this paper, we present the results of the search for very high redshift 
galaxies ($z\sim4.5$ ), using the same technique, but with an original 
instrumental setup : we have designed multi band filters, with three 
narrow ($\mathrm{FWHM} \simeq 200$~\AA) bands, placed so that they
avoid the most prominent night sky emission lines. This disposition 
and the very high transmission (above 95\%) of the filters
compensate the narrowness of the bands, and reduces by a factor of three the
amount of data to reduce, as well as telescope observing time. 

Observations are described in Sect.~\ref{data}, and our detections are 
discussed in Sect.~\ref{res}.
\section{\label{data}Observations and data reduction}
We have observed the field \#F of the Calar Alto Deep
Imaging Survey (Hippelein et al. 1995), centered at
position 13h 47m 42.1s $+05\deg$ 37\arcmin 35.6\arcsec\
(J2000). The CADIS fields have been selected for being empty of
bright stars or galaxies. Observations were carried at the ESO 3.6m 
Telescope at la Silla between the 2nd and the 5th of April 1997, with EFOSC. 
The Tektronik CCD has a 0.61\arcsec\ pixel size. 

\begin{figure}
\centerline{\psfig{figure=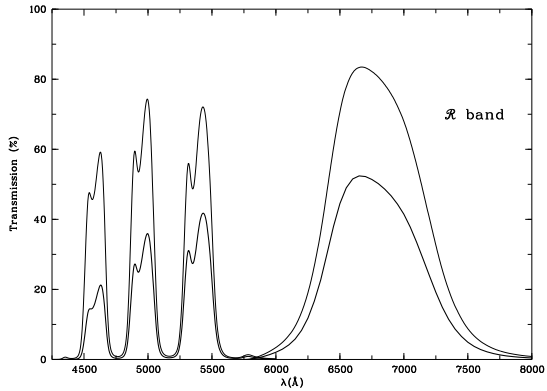,width=8.0cm,clip=t,angle=-90}}
\caption{\label{filter} Transmission curves of the optical device alone 
(multi-band filter and prisms, thin curve), and of the overall setup (optical
device, EFOSC and CCD, thick curve)}
\end{figure}

We have taken two exposures of the field : one in the standard $R$ filter, so
that we could locate precisely the objects and measure their magnitude, and
one with the 3-band filter, to detect their Lyman edge.
The full instrumental setup and properties are described in Boulesteix et
al. (1997). In short, the three bands have been designed to avoid the most 
prominent night sky emission lines (5700~\AA, 6003~\AA), so that the
background noise is significantly reduced.
The bandpass of the multi-band filter is shown on the left part of 
Fig.~\ref{filter}, while the right part of this figure gives the transmission
of the $R$ filter. We both show the transmission of the optical device
(filter + prisms or $R$ filter), and the curve obtained when including the 
transmissions of the prisms, the multi-band filter and the broad band filter 
that is used to avoid transmission outside the desired wavelength
range (due to residual transmission of the multi band filter below 4450~\AA\ 
and above 5600~\AA), and EFOSC and CCD efficiencies. We have checked that the
optical device does not induce any distorsion from center to edge of the field,
neither in photometry nor in the relative locations of the three images of a
given object. 
 
The separation of the images in the 3 bands of our filter, 2.7\arcsec, has 
been calculated to avoid crowding with the closest objects on the sky and is 
close to the limit
resolution of the telescope and seeing. It assures an almost gaussian
image without overlap of the images. The average projected density 
of galaxies with apparent magnitude below $\mr=25$ is given by the integration
of the differential
magnitude-number counts of field galaxies given by e.g. Le~Brun et al. (1993): 
\begin{equation}
\label{dnds}
{\partial^2 N\over \partial S \partial m} = 10^{\left(0.366\pm 0.01\right)\mr -
  (4.24\pm0.31)} \mathrm{deg}^{-2},
\end{equation}
 in very good agreement with those obtained by  Metcalfe et al. (1991), Smail 
et al. (1995) or Driver et al. (1994). 
This gives an average distance between galaxies of about 10\arcsec. Thus, with 
the two extreme images separated by 5.8\arcsec, we avoid crowding in  a large 
majority of cases. 

The flat fields were made by averaging all the exposure in the same filter 
($R$ or multi-band), and the final image was reconstructed by using the
shift-and-add procedure. The total exposure time is thus 2 hours in the $R$
band and 10 hours  in the multi-band filter and the final field of view,
reduced by vignetting and shift-and-add procedure, is 4\arcmin50\arcsec\
square. Detection of the objects on the $R$ exposure was carried using the
Sextractor software (Bertin \& Arnouts 1996). Thus, knowing the position of
all
the objects on the field, we have performed aperture magnitude, in a
2\arcsec\ box, of the three images of each object on the 3-band filter
exposure, at the expected location of each image. We have verified on objects
for which the three images were detected that our expectation of the location
of the images was correct. Photometric and flux
calibrations were made in the $R$ band using the field Ser~1 (Tyson 1988), and
the calibration of the multi-band filter exposures was made using the
spectrophotometric standard star Kopff~27 (Stone 1977). 

For our search of very high redshift galaxies, we have selected objects with
apparent magnitude in the range $23.5\le \mr\le 25$. The lower limit
corresponds to objects for which the three images are severely blended (due
to the large 
pixel size), and the upper limit to the limiting magnitude of the $R$ image. 
Thereafter, we have selected, among the 352 objects in this magnitude range,
those for which the image quality was good enough to perform photometry of the 
three images. We note however that blending of objects with each other was
in our case less a crucial problem than the blending between the three
images of an individual object due to the large pixel size of the 
EFOSC CCD. We then come up with a list of 88 objects, for which we could
perform 
efficient photometry. We have thus verified, using Kolmogorov-Smirnov test,
that this selection did not introduce any bias in the spatial distribution of
the objects, and in fact neither the abscissa nor ordinate distribution of
the selected objects do deviate significantly from an uniform distribution.

Among these objects, four show image neither in the intermediate
nor in the bluer band,
and are thus candidates for being galaxies in the redshift range 
$4.55\le z\le 4.76$,
and four more show only two images in the two redder bands, thus being 
potentially at redshifts in the range $4.12 \le z \le 4.32$. 
\begin{table}
\caption{\label{cand} Characteristics of the candidates}
\begin{tabular}{rrrrrr}
\hline\noalign{\smallskip}
 Object  & $\mr$ & F(6800) & F(5400) & F(4950) &  F(4590)  \\
         &       &
\multicolumn{4}{c}{($10^{-19}$~erg~cm$^{-2}$~s$^{-1}$~\AA $^{-1}$)} \\
\hline\noalign{\smallskip} 
\multicolumn{6}{c}{High redshift candidates ($4.55\le z\le 4.76$)}\\
H1 & 24.5 & $2.71\pm 0.8$ & $4.03\pm 0.63$ & $\le 1.10$ & $\le 1.50$ \\
H2 & 23.8 & $5.18\pm 1.0$ & $4.56\pm 0.63$ & $\le 1.10$ & $\le 1.50$ \\
H3 & 24.3 & $3.26\pm 0.6$ & $4.54\pm 0.53$ & $\le 1.01$ & $\le 1.37$ \\
H4 & 24.1 & $3.92\pm 0.8$ & $6.68\pm 0.63$ & $\le 1.10$ & $\le 1.50$ \\
\hline\noalign{\smallskip}
\hline\noalign{\smallskip}
\multicolumn{6}{c}{Low redshift candidates ($4.12 \le z \le 4.32$)}\\
L1 & 23.8 & $5.17\pm 0.6$ & $4.77\pm 0.63$ & $5.15\pm 1.10$ & $\le 1.50$ \\
L2 & 24.2 & $3.58\pm 0.5$ & $6.68\pm 0.53$ & $4.97\pm 0.91$ & $\le 1.25$ \\
L3 & 24.3 & $3.26\pm 0.6$ & $4.77\pm 0.53$ & $4.42\pm 0.91$ & $\le 1.25$ \\
L4 & 23.9 & $4.72\pm 1.0$ & $2.76\pm 0.47$ & $3.50\pm 0.83$ & $\le 1.13$ \\
\hline\noalign{\smallskip} 
\hline\noalign{\smallskip} 
\multicolumn{6}{c}{Emission line object}\\
E1 & 23.7 & 5.89 & 4.63$\pm 0.6$ &14.82$\pm 1.1$ & 7.37$\pm 1.5$ \\
\hline\noalign{\smallskip} 
\end{tabular}
\end{table} 
\begin{figure}
\centerline{\psfig{figure=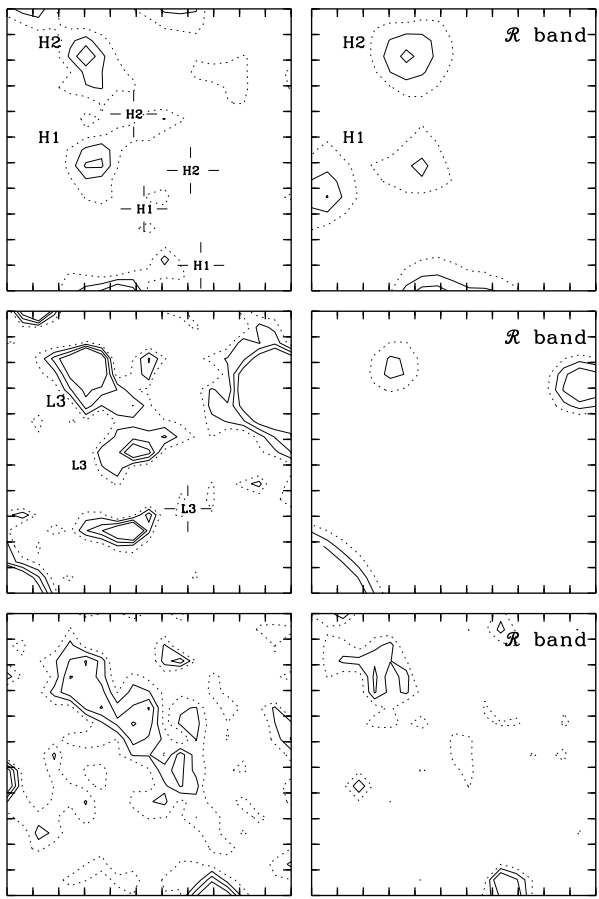,width=8cm,clip=t}}
\caption{\label{cands} Images of different types of objects. Upper panel 
shows the two close higher redshift candidates H1 and H2, intermediate panel 
shows the $4.13\le z\le 4.38$ candidate L3, and the lower panel shows an 
object of comparable magnitude which shows the three band in emission, for 
comparison sake. The dispersion direction goes from top-left to
down-right, with decreasing wavelength. The symbols at the center of
the crosses indicate the expected location of the blue image(s)
}
\end{figure}
\section{\label{res}Results and discussion}
We present in Table~\ref{cand} the characteristics of the candidates, for both
redshift ranges. The R magnitude is given in Column~2, and the calibrated 
fluxes (or the limits) in the Columns 3 to 6. 
\subsection{\label{high_z_sec}The higher redshift candidates}
Upper panel of Fig.~\ref{cands} presents the four  galaxies candidates at 
redshifts $4.55 \le z\le 4.83$, both in R band and in the multi-band filter 
image. For these four objects, the flux in the intermediate band is at least 
3.7 times smaller than
expected, if the flux had a flat distribution. 

\begin{figure}
\centerline{\psfig{figure=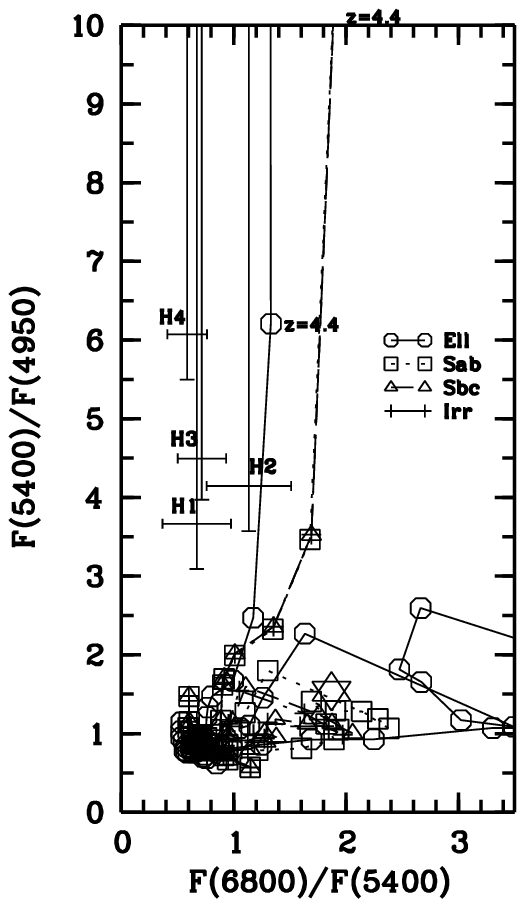,width=4cm,clip=t}\psfig{figure=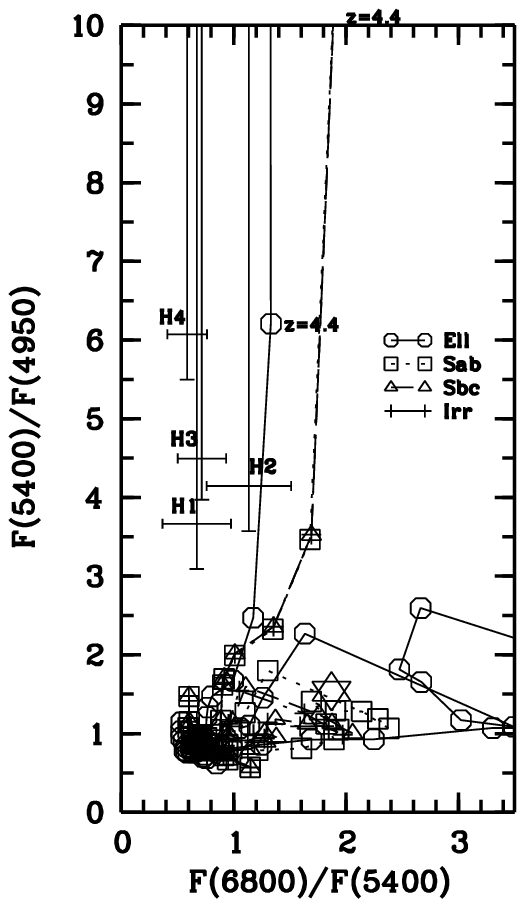,width=4cm,clip=t}}
\caption{\label{col_hz} Color-color diagram for the highest redshift
candidates. $F(6800)$ denotes the fux in the $R$ band, while
$F(5400)$ and $F(4950)$ are the fluxes in the two redder bands of the
multiband filter. The location of our four candidates, with the
$3\sigma$ error bars, is also shown. The left diagram is for a high value of
the intergalactic transmission, right plot for an average transmission. The
star-like symbol show the colors of an M type star
}
\end{figure}

The main concern in such a photometric study is the possible
contamination by lower redshift objects, whose spectra show strong
features (e.g. ellipticals), that can produce the same effects that
very high redshift galaxies. Thus, in the same way as Steidel et al.
(1995), we have simulated the evolutions of the colors of different
standard
types of galaxies for all redshifts between 0 and 5. The results for the
colors relevant for the high redshift objects (namely
$F(5400)/F(4950)$ vs $F(6800)/F(5400)$) are given in
Fig.~\ref{col_hz}. We have included in the simulations the effect of
the absorption by the intergalactic medium (the so-called \lya\
forest), following the prescription of Madau (1995), specifically his
Figure 3. As can be seen from this figure, the Universe is transparent
above the redshifted wavelength of the \lya\ line, and totally opaque
below the redshifted Lyman limit. The absorption between these values
is very uncertain, due to the statistical fluctuation of the number of
absorbers toward a particular sightline. We thus have performed
several simulations with various values and form of the intergalactic
absorption. We show two of them, corresponding to the average and highest
transmission curve. As can be seen, the only effect of a stronger \lya\
forest absorption is to make all objects redder in $F(6800)/F(5400)$,
but does not modify by a significant amount the $F(5400)/F(4950)$, since both
bands are 
affected in the same proportion. Thus, the value of $F(5400)/F(4950)$
alone can be used to detect very high redshift galaxy candidates,
since a galaxy of any type at lower redshift cannot reproduce the
observed colors. As well, faint M stars do not have colors compatible with
those of high redshift galaxies.

Thus, in the hypothesis that these objects are at redshift above 4.5, objects
H1 and H2 are located 3.6\arcsec\ apart from each other (see upper panel of 
Fig.~\ref{cands}).
This represents a projected separation of $50\h50$~kpc, and we would thus see
a close pair of primordial galaxies. 

\begin{figure}
\centerline{\psfig{figure=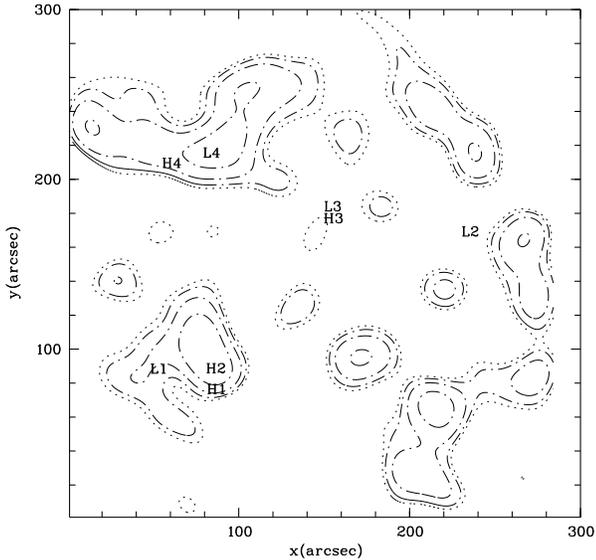,width=8cm,clip=t,angle=-90}}
\caption{\label{dens} Galaxy density map of the CADIS field. The
contours are 2 (dotted line), 4,8 and $16\sigma$ (solid lines) above
the mean value (poissonian statistics). The location of our 8
candidates is also shown. 
}
\end{figure}

We have drawn
the galaxy density map in the field. This is shown on Fig.~\ref{dens},
together with the location of all our high redshift candidates. As can
be seen, the pair H1/H2 is located on the edge of an overdensity,
which is significant at $16\sigma$ level. No other high
redshift candidate is detected in the same area of the field, due to
useless photometry, and they could belong to a very high redshift
structure.

In the hypothesis that these objects are at redshifts around 4.6, and 
correcting the fact that we considered only 88 objects among the 352 
that we detect in the apparent magnitude range $[23.5,25]$, we come to 
a projected density of 0.7 object per square arcminute. This value is 
very close to the one derived by Steidel et al. (1995) at redshifts 
around 3.2, thus involving no strong evolution of the projected 
density of galaxies between $z=4.5$ and $z=3.2$. We note however that, due to
cosmological evolution, this leads to strong change in the physical density, 
but that Mirales \& Pello (1998) have recently shown that the density of high 
redshft objects could be as high as 50 times higher than the one derived by 
Steidel et al. (1995)
\subsection{\label{low_z_sec}The lower redshift candidates}
\begin{figure}
\centerline{\psfig{figure=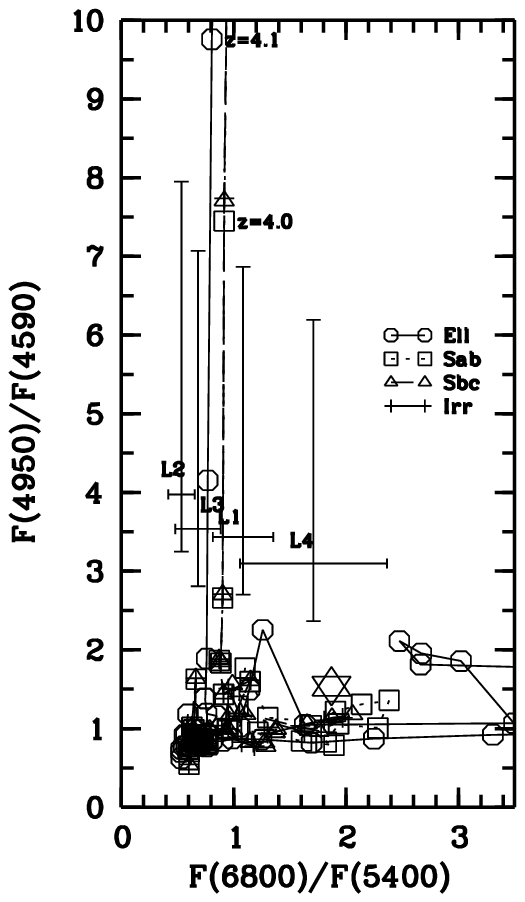,width=4cm,clip=t},\psfig{figure=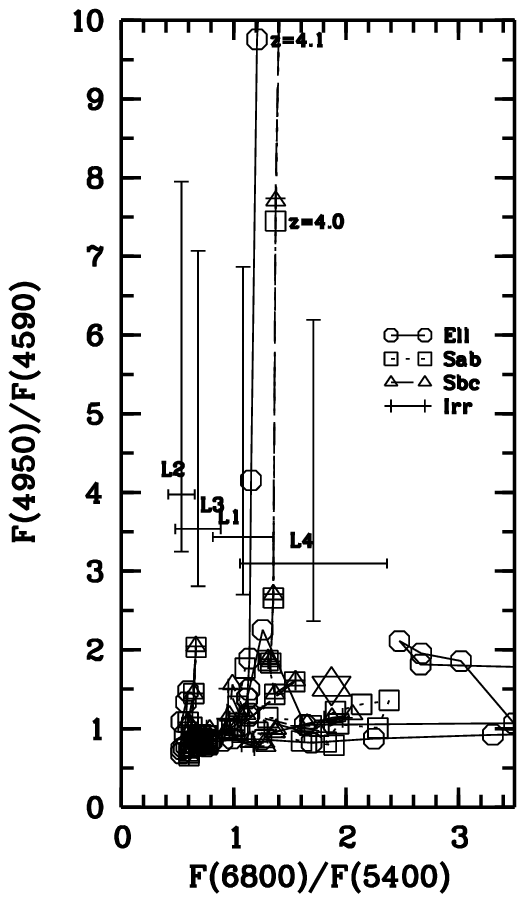,width=4cm,clip=t}}
\caption{\label{col_lz} Color-color diagram for the lower redshift
candidates. $F(6800)$ denotes the fux in the $R$ band, while
$F(4950)$ and $F(4590)$ are the fluxes in the two bluer bands of the
multiband filter. The location of our four candidates, with the
$3\sigma$ error bars are also shown.Left diagram is for a high intergalactic
transmission, right diagram for average value of the transmission. The
star-like symbol shows the colors of an M type star
}
\end{figure}

Intermediate panel of Fig.~\ref{cands} shows the images of object L3, which is
likely to be a galaxy with redshift in the range $4.12 \le z \le
4.32$. Again the color-color plot 
($F(4950)/F(4590)$ vs $F(6800)/F(5400)$ for this redshift range, Fig.~\ref{col_lz}) shows that no 
lower redshift objects can reproduce the observed colors, and that the
$F(4950)/F(4590)$ value alone can discriminate high redshift objects.
The only possible exception is object L4, which has colors compatible
at slightly more than $3\sigma$ with lower redshift ellipticals at $z=0.2$ or
$z=0.7$.

\subsubsection{Peculiar objects}
This experimental setup also has the capability to bring out objects
with other kinds of spectral feature, in particular emission lines. In
our sample of faint galaxies, we have found one such object, labeled
E1 in table~\ref{cand}. The flux in the intermediate band  is about 3
times greater than in other bands (including the R band). It is thus
likely that this object shows a strong emission line in the wavelength
range 4855--5065~\AA . Unfortunately, there is no way to distinguish
between the various lines, and this object could either be a \lya\
emitting object at $z\sim 3.1$, or a local dwarf object ($\Mr \sim
-18$) at $z \sim 0.33$ which shows strong O\,{\sc ii} emission. 

\acknowledgements{This experiment has been made possible thanks to a strong
support of the IGRAP. We also would like to thank E. Pelletier and G.
Albrand for the realization of the very high transmission multi-band
filters, as well as A. Baranne for doing all optical computations, and
B. di Biagio and L. Castinel for the realization of the mechanical
structure. We also thank H.H. Hippelein for kindly providing us with
the CADIS field position and finding charts, A. Gilliotte for his
essential help at the ESO-La Silla Optical Laboratory during the
observations, and H. Plana and P. Ambrocio-Cruz for their help during the
observations. We thank V. Afanasiev, S. Dodonov and H. Hippelein for fruitful 
discussion. V. Le Brun is supported by a grant from the "Soci\'et\'e
de Secours des Amis des Sciences" of the Acad\'emie des Sciences de
Paris}


\begin{thebibliography}{}
\bibitem[]{} Bertin E., Arnouts S., 1996, A\&AS 117, 393
\bibitem[]{} Boulesteix J. et al., 1997, in preparation
\bibitem[]{} Bruzual G., Charlot S., 1993, ApJ, 405, 538
\bibitem[]{} Charlot S., Fall S,., 1993, ApJ, 405, 538.
\bibitem[]{} De Propris R., Pritchet C., Hartwick F., Hickson P. 1993, AJ, 105, 1243
\bibitem[]{} Driver S.P., Phillipps S., Davies J.I., Morgan I., Disney M.J.,
1995, MNRAS 266, 155
\bibitem[]{} Franx M., Illingworth G.D., Kelson D.D., van Dokkum P.G., Tran K.-V.,
1997, ApJ 485, L75, astro-ph/9704090
\bibitem[]{} Hammer F., Flores H., Lilly S. J., et al., 1997, ApJ 481, 49
\bibitem[]{} Hippelein H., Meisenheimer K., Thommes E., Fockenbrock R.,
R\"{o}ser H., 1995, "New light on Galaxy Evolution", IAU Symp 171, 
Heidelberg, June 26-30, 1995.
\bibitem[]{} Hu, E., In S. D'Odorico, A. Fontana and E. Giallongo (Eds), ``The
       Young Universe: Galaxy Formation and Evolution at Intermediate and High Redshift'', ASP Conf. Series, astro-oh/9801170
\bibitem[]{} Koo D., Kron R., 1980, PASP, 92, 537
\bibitem[]{} Le~Brun V., Bergeron J., Boiss\'e P., Christian C., 1993, A\&A 
  279, 31
\bibitem[]{} Madau P., 1995, ApJ 441, 18
\bibitem[]{} Metcalfe N., Fong R., Shanks T., Jones L.R., 1991, MNRAS 249, 498
\bibitem[]{} Miralles J.-M., Pello R., 1998, submitted to ApJ, astro-ph/9801062
\bibitem[]{} Parkes I., Collins C., Joseph R., 1994, MNRAS, 266, 983
\bibitem[]{} Schneider P., Schmidt M., Gunn J.,1991, AJ, 102, 837
\bibitem[]{} Smail I., Hogg D.W.,  Yan L., Cohen J., 1995, ApJ 449, L105
\bibitem[]{} Steidel C.C., Hamilton D., 1992, AJ 104, 941
\bibitem[]{} Steidel C.C., Hamilton D., 1993, AJ 105, 2017
\bibitem[]{} Steidel C.C., Pettini M., Hamilton D., 1995, AJ 110, 2519
\bibitem[]{} Stone R.P.S., 1977, ApJ 218, 767
\bibitem[]{} Thompson D., Djorgovski S., 1995, AJ, 110, 982
\bibitem[]{} Thompson D., Djorgovski S., Trauger J., 1995, AJ, 110, 963
\end{thebibliography}
\end{document}